\documentclass[12pt]{article}
\usepackage{graphicx} 
\usepackage[margin=1.1in]{geometry}
\usepackage{subcaption}
\usepackage{amssymb}
\usepackage{amsthm}
\usepackage{amsmath}
\usepackage{xcolor}
\usepackage{url}

\NewCommandCopy{\proofqedsymbol}{\qedsymbol}

\AtBeginEnvironment{proof}{\renewcommand{\qedsymbol}{\proofqedsymbol}}

\theoremstyle{definition}

\newtheorem{example}{Example}

\title{Grammatical structures in mathematics:\\ a personal view}
\author{Tess O'Brien}
\date{July 2025}

\begin{document}

\maketitle
\begin{abstract}
    The ability to read, write, and speak mathematics is critical to students becoming comfortable with statistical models and skills. Faster development of those skills may act as encouragement to further engage with the discipline. Vocabulary has been the focus of scholarship in existing literature on the linguistics of mathematics and statistics but there are structures such as grammar that go beyond the content of words and symbols. Here I introduce ideas for grammar structures through a sequence of examples.
\end{abstract}

\paragraph{Acknowledgments}
Thanks to Laura Le for directions in the literature and a keen eye on drafts, Jacqui Ramagge for pointers on the adjective-noun description in the work of Herb and Ken Gross, Alberto Nettel-Aguirre and Elinor Jones for advice on drafts. Thanks also to the reviewers for their feedback and expertise.

\section*{Introduction}
When we teach statistics, we do so in the language of mathematics. When we construct models, communicate them, use them to make predictions, we do so in the language of mathematics. This use of mathematical language in statistics is distinct from the language of statistics itself as centred on careful framing of problems and analysis. Our students struggle to learn mathematical language because it is hard, and the linguistic structures are opaque even to the fluent. 

Explicit grammar teaching is part of both first and second language instruction more broadly. In the first language case, the justification is both based in literacy development as in Systemic Functional Linguistics research such as \cite{macarthur2008handbook}, and more recently in the development of reasoning and critical thinking skills in \cite{van2021conceptual} and \cite{hudson2004education}. In second language teaching, explicit grammar instruction is used to help with meaning making in context as well as learning by comparison to the first language \cite{pawlak2021teaching}. By making grammar in mathematics explicit we may be able to help students develop confidence and competence in using mathematical language to express their ideas, just as explicit teaching of grammar helps in other language teaching settings. To make grammatical structures in mathematics explicit, we first need to know what they are.

That mathematics is a language is not a new idea. In many ways the revolution in proof theory during the 19th and into the 20th century was about the language of argument structure. There has been work since then in linguistic discourse theory including \cite{o2008mathematical} from the Systemic Functional Linguistics perspective, which does not address the syntax-level structure I wish to discuss. In mathematics education research, works such as \cite{bruun2015language} and \cite{powell2021assessment} have predominantly addressed the vocabulary used to pose mathematical problems in school learners. From statistics education, \cite{rangecroft2002language}, \cite{kaplan2010lexical}, and \cite{dunn2016learning} addresses the lexicon of statistical English (vocabulary) and its interaction with ordinary and mathematical English for comparison. However, the lexicon is only part of a bigger picture of language structure, as grammatical structure and processes are also key to language understanding \cite{van2021conceptual}. Recent developments in Systemic Functional Linguistics as applied to mathematics within physics education \cite{doran2017discourse} and Piaget's ideas about mechanisms of abstraction \cite{dawkins2024piaget} have opened up some useful avenues for thinking about more general linguistic structures in mathematics, advancing ideas in \cite{obrien2023mathslanguage}.

In this article I will introduce and apply ideas about grammatical structures and processes in mathematics to some specific statistical examples. I use English as the underlying language, as while there may be similar structures in other language, I am not fluent in them to say. I start with a simple mathematical example, then extend those ideas to more complicated situations that arise in statistics that demonstrate the same basic ideas of syntactic structure in mathematical representations.

\section*{Example: Linear Function}
The first example comes from mathematics more than statistics, to set up the linguistic tools we will use later. I introduce some terminology for word classes -- types of words defined by their function such as nouns and verbs -- through the example and then talk about a grammatical process where words move between classes based on use.
\begin{example}
    Let $f$ be a function from $\mathbb{R}$ to $\mathbb{R}$ given by
    \begin{equation}\label{eqn_2x1}
        f(x) = 2x+1.
    \end{equation}
    As an example $f(2) = 5$.
\end{example}
When we first use $f$ it is as a name, a noun. The second occurrence -- $f(x)$ -- is as a modifier on the argument $x$, which produces a noun group or noun phrase: a collection of lexical items that together behave like a noun. Within the group, the noun is $x$, and $f$ modifies the noun in the same way as `painted' does in `the painted chair'. In the case of $f(x)$, we have in fact $f$-ed $x$. I call this the \emph{modifier} form of $f$, a more general term for the class of words that adjectives belong to because it has a similar role. The process of $f$-ing a specific $x$ is what we do when we calculate the result. The third occurrence, $f(2)$, is another noun phrase.

Our introduction of $f$ as a noun is a \textit{performative} linguistic act: we are doing a speech act that itself causes something to happen. Think of a promise or a declaration by an authority figure or a variable initialised in code. `Let $f$ be a function' instantiates the mathematical object called $f$ in the minds of the readers and ascribes to it the structures of a function. We do this when we announce random variables too, for example.

Words moving between classes is extremely common in mathematics because we tend to start by using a mathematical object to modify a noun or as an operation, then refer to it as a noun to add information. In linguistics, a word becoming a noun is \textit{nominalisation} and happens commonly to verbs and adjectives. \cite{obrien2023mathslanguage} points to numbers as adjectives in a construction like counting objects in the world where the number describes the noun. Five chairs, three plates, 15 people. The technically correct term is \emph{quantifier}: words that modify the quantity of a noun and includes approximate terms like `some' or `many'. Numbers get nominalised when we start talking about them as mathematical objects in the abstract -- `5 chairs' to `5'. In Example 1 we are moving between word classes for the function but in the reverse direction to becoming a noun. We start with naming the object $f$ and assigning it the properties of being a function, then use $f$ to modify a noun when we apply it to $x$. 

More generally, we can associate the grammatical process of nominalisation in mathematics with moving between levels of abstraction. The mathematics education literature on abstraction is extensive and typically connected to the process of learning mathematics through the work of Jean Piaget. I think the most relevant ideas are those in the Piaget process of reflective abstraction \cite{dawkins2024piaget}. If we consider the use of a mathematical object (quantifier form of the number, modifier $f$ for example) to be a lower level of abstraction compared to what that thing becomes post-nominalisation (numbers or functions as nouns), we can align the grammatical process with learning the next level in mathematical structure. Once a number has been nominalised, we start talking about its properties. For functions, when we start treating them as objects rather than modifiers to arguments we introduce concepts like domain and range, continuity, injectivity and surjectivity. Classes of functions such as polynomials require nominalisation because we define them as all functions with a specific property. In a similar way, we nominalise actions into exercises -- running, swimming etc. -- and can classify those exercises based on characteristics.

In Equation \eqref{eqn_2x1} we also see an example of 2 becoming un-nominalised, returning to quantifier form. The construction of $2x$ as a noun group uses 2 as a quantifier describing the noun $x$. For languages other than English that use a morphological structure for quantifiers, this may be more directly compared to a quantifier prefix 2 and a noun root $x$ to give the compound $2x$. Herb and Ken Gross are credited with an adjective-noun (technically quantifier-noun) description of these algebraic objects many years ago (the adjective-noun algebra description has inspired many people as evidenced by the material on the now-defunct website \texttt{adjectivenounmath.com} \cite{adjectivenoun2012}) but I disagree with their claim that arithmetic alone without variables (such as) 
\begin{equation}\label{eqn_2_plus_1}
    2+1=3
\end{equation}
has a hidden common noun that the numbers are attached to as quantifiers. Instead I think that in Equation \eqref{eqn_2_plus_1}, 1, 2, and 3 are present as \emph{nouns}, and it is plausible that one of the difficulties of introducing the quantifier-noun algebraic construction is undoing the nominalisation process. For comparison, in Equation \eqref{eqn_2x1} we go from $2\times x$ where both 2 and $x$ are present as nouns to $2x$ -- the quantifier-noun group construction that has been transformed grammatically and mathematically by the now hidden multiplication process. We still have 1 appearing as a noun though, so the student has to hold both grammatical forms in play simultaneously.

In the first example we have introduced some linguistic terminology in relation to a simple function definition. From here we are going to move into a far more complicated function that is more directly relevant to teaching statistics -- the normal distribution cumulative distribution function.

\section*{Example: Probability Density Function}
As a second example, let us consider the normal distribution density function. We turn the grammatical structure analysis to parts of the equation in greater depth than we did in the first example. Students may benefit treating a complex equation as a pictogram with grammatical structure, just as teaching the orthographic structure of pictograms in traditional Chinese assists second language learning \cite{tong2015cracking, hong2016effect}.

\begin{example}
    The probability density function for the normal distribution with mean $\mu$ and variance $\sigma^2$ is
    \begin{equation}\label{eqn_PDF}
        f(x) = \frac{1}{\sqrt{2 \pi \sigma^2}}\exp \bigg(-\frac{(x-\mu)^2}{2\sigma^2} \bigg).
    \end{equation}
\end{example}

As in the first example, our naming of the cumulative distribution function $f(x)$ is instantiation as a noun group. I read $=$ as `is equal to', which combines the verb `is' and the prepositional phrase `equal to', where the prepositional phrase is used to express a relationship, in this case mathematical equality between the two sides.

On the right hand side of Equation \eqref{eqn_PDF}, the first part of the pictogram is $\frac{1}{\sqrt{2 \pi \sigma^2}}$. On top of the fraction is the noun 1, with the fraction line acting as the verb of division. The denominator $\sqrt{2 \pi \sigma^2}$ which is a noun group. The square root function is a modifier on the argument $2 \pi \sigma^2$, as is the power on $\sigma$. Multiplication is implicit between the nouns $2$, $\pi$ and $\sigma^2$, as multiplication combined them into a noun group.

Next we have the exponential function and its argument, which is combined with the first part through implicit multiplication. As with the other functions, $\exp$ can be understood as a modifier on the noun group argument. The argument in this case is quite complicated. We start with $-$ as a negative modifier on the fraction. The numerator has a power modifying the phrase $x-\mu$, which is a second use of $-$ but this time as a verb combining the nouns $x$ and $\mu$. The dual use of $-$ as an arithmetic operation and a valence modifier indicating a negative number, the first of which takes two arguments and the second one, can be unhelpful. The fraction line represents division as a verb, and the denominator is another noun phrase constructed through multiplication of noun $2$ and noun phrase $\sigma^2$. 

In Equation \eqref{eqn_PDF} we have several ways of relating functions and operations to arguments. We suppress notating multiplication, instead using it to construct noun groups. Subtraction is explicit, with arguments positioned to the left and right. Division is also explicit, but arguments are above and below. The square root function has a line over the arguments to represent where that noun group ends, while the exponential function and $(x-\mu)^2$ squaring operation use parentheses instead. In the absence of parentheses, $\sigma^2$ is interpreted as only applying to $\sigma$. Just as lexical ambiguity confuses students in \cite{kaplan2010lexical}, multiple ways of expressing relationships between an operation and its arguments may raise a barrier to entry.

In the normal distribution PDF we see much more complicated grammatical structure through the construction of Equation \eqref{eqn_PDF} when compared to Equation \eqref{eqn_2x1}. Some similar relationships are at play such as constructing noun groups through implied multiplication, but the introduction of more functions acting as verbs on their arguments makes the pictogram harder to deconstruct.

In the next example we move to multiple linear regression, which behaves more like Equation \eqref{eqn_2x1} in the underlying structure (as a linear combination), but has additional features such as the combination of multiple terms.

\section*{Example: Multiple Linear Regression}
In this example we will present the regression equation in two forms: first with matrix notation and then with explicit arithmetic. The matrix form relies on changing the definition of arithmetic operations to be matrix operations, which can be ambiguous, but uses the same grammatical structure as Example 1.

\begin{example}
    Consider data with variables $Y$, $X_1$, and $X_2$, transformed variable $X_1^2$, and interaction term $X_1X_2$ under the linear model
    \begin{equation}\label{eqn_MatrixLinReg}
        Y = X\beta + \epsilon.
    \end{equation}
    We can also express Equation \eqref{eqn_MatrixLinReg} with direct reference to data as
    \begin{equation}\label{eqn_ArithLinReg}
        y_i = \beta_0 + \beta_1 x_{1i} + \beta_2 x_{2i} + \beta_3 x_{1i}^2 + \beta_4 x_{1i} x_{2i} + e_i.
    \end{equation}
\end{example}

The expression in Equation \eqref{eqn_MatrixLinReg} has a near-identical grammatical construction to Equation \eqref{eqn_2x1}: on the left side a noun $Y$, on the right hand a noun group $X\beta$ and noun $\epsilon$ combined by the verb $+$. However, we have replaced the operations with matrix rather than number arithmetic. In the matrix form, it is less appropriate to read $\beta$ as a modifier on $X$, but rather as a noun in its own right. The implicit change in operations to matrix arithmetic can trip students up because we use the same arithmetic symbols and grammatical construction -- we introduce lexical ambiguity through multiply defined symbols. We can still read $Y$, $X$, and $\epsilon$ as nouns, and $+$ as a verb, but there is a considerably greater amount of information tied up in the symbols compared to Equation \eqref{eqn_2x1} because the data contained in $X$ is no longer a single number, and the operations of matrix arithmetic are combinatorial number arithmetic rather than a single instance thereof. The high density of information of Equation \eqref{eqn_MatrixLinReg} is precisely why we don't require any more symbols to express the relationship, but comes at the cost of making the matrix operation structure implicit. 

Equation \eqref{eqn_ArithLinReg} is a more numerically explicit expression of the same relationship as Equation \eqref{eqn_MatrixLinReg}, but in lowering the amount of information contained in each symbol, we have to lean on many more symbols being present to encode the linear structure directly. By writing out the arithmetic we can talk much more directly about the variables, and have easier tracking of transformations and interactions. We can leverage the grammatical construction of quantifier-noun groups from the first example too. The terms in Equation \eqref{eqn_ArithLinReg} like $\beta_1 x_{1i}$ are such constructions where $\beta_1$ is the quantifier, and the noun groups get combined with the verb $+$. We still have $\beta_0$ as a noun on its own, analogous to the presence of the constant 1 in Equation \eqref{eqn_2x1}.

Equations \eqref{eqn_MatrixLinReg} and \eqref{eqn_ArithLinReg}, combined with the initial naming of the variables and statement that they have a linear relationship, gives us three different representations of the same underlying structure. There are differing levels of abstraction depending on how explicit the number arithmetic is made. The term `linear model' is the most abstract, and is the nominalisation of the overall relationship. I think part of mathematical fluency is the ability to move between different representations of the same object depending on which highlights relevant structure at the time, a simple example of which would be going back and forth with summation notation such as
\[ 1+2+3+4+\ldots +n = \sum_{i=1}^n i.
\]

We can see the linguistic choices in notation in this example, and how those relate to linguistic structure. We also get to see different levels of abstraction directly linked to how explicit or implicit the mathematical structure is for the particular model. As we use linear regressions to represent systems of interest in the world, it's worth looking at how that relationship works from the perspective of language.

\section*{In the classroom}
The core argument of this article is that making linguistic structure clear to students may help them leverage their existing language skills to understand mathematical objects and representations. Existing skills may include formal grammar training in schools, which makes it easier to introduce terminology such as nouns and verbs, but in the absence of language structure education it may still be possible to use direct comparison to phrases with similar construction. For example $2x$ is comparable to `two apples', which I have observed in mathematics classrooms at the high school level but without the explicit intent of leveraging grammatical structure.

As statisticians, we use mathematical objects as models, and rely on mathematical language to describe them, which can be a barrier to statistical users and students who do not have the opportunity to develop good mathematical fluency. Understanding language structure in mathematical expressions and making them explicit may help overcome such a barrier. How mathematical language is used in the specific context of statistics may be different to, for example, theoretical mathematics.

I have explicitly taught these grammatical structures in my classes and found it useful, but there is no literature testing the results as the linguistic structures I am discussing are not part of existing language theory. However, I can offer some suggestions for how to introduce the grammatical structure I discuss. 

We can use ideas like nouns and verbs to interpret what is going on in a symbolic expression, and second that there are grammatical processes at play that move specific symbols between word classes in order to convey different information. Demonstrating this with explicit examples is key because it is difficult to conceptualise in the abstract. As seen in Example 1, examples do not need to be very complicated to get started and I have found them particularly useful for non-mathematics students who are not exposed to enough mathematics to develop fluency otherwise. Those ideas are applicable to more complicated cases such as Example 2, where there are nested operations and functions at play. 

In Example 3, we have different representations of the same mathematical object, relying on more or less explicit expressions of the linear relationship. The most abstract is the nominalised form `linear model', while the least abstract gives the numerical arithmetic as a combination of noun groups.

\section*{Discussion and Conclusions}
In this article I have shown a few linguistic structures that I believe are at play when we use mathematical language in statistics. These structures exist at multiple levels, from the relationships between symbols in an equation to how we refer to mathematical objects in text around equations. Making these structures explicit may help students leverage their existing language skills including the basic grammar they learn in school, and help them to see the process of learning to work with the mathematical objects we use as a process of learning the language of mathematics. Evidence from second language learning indicates that explicit teaching of grammar is helpful \cite{hudson2004education, pawlak2021teaching}.

At the level of static symbols we have pictograms such as Equation \eqref{eqn_PDF}, where a complicated web of syntactic relationships -- many implicit and based on relative position -- presents a major barrier to learners. Equipping students with the tools to dissect these expressions as a construction in \textit{language} may help to relieve the anxiety of being confronted with such a mess. Equation \eqref{eqn_ArithLinReg} leverages fewer grammatical tools, but relies on the recognition of noun groups connected up by verbs in order to separate out the pieces.

Grammatical processes come into play in more dynamic situations across the three examples. Moving a particular symbol between word classes is key to how we first introduce, and then use, a mathematical object such as a function. Students really struggle to recognise that is what they are doing when they write and use equations, and can get confused about the role symbols play as a result. The reverse process of nominalisation plays a key role in how students are first introduced to many different mathematical objects, and I suspect is a key barrier to developing mathematical fluency.

The ideas presented in this article are based on \emph{English} mathematical reading specifically, but many of the same structures are likely transferable, such as nouns and quantifiers. In a mixed language classroom, inviting students to map out grammatical structures across languages may reinforce learning, but would take time and effort that is rarely available to teachers.

There is much work yet to be done on linguistic structure in mathematics and statistics. The existing research from Systemic Functional Linguistics has operated at the discourse level in \cite{o2008mathematical} and \cite{doran2017discourse}, while research from mathematics and statistics education has focused on lexicon (\emph{e.g.} \cite{rangecroft2002language}, \cite{dunn2016learning}, \cite{kaplan2010lexical}) and does not address grammatical structure and processes. However, in the absence of a robust theory we can leverage observed structures to make it easier for students to be aware of what they are doing, and make better choices about how they use mathematical expressions to represent statistical models. 

Due to the absence of existing work, there is a fertile field of research available in the linguistics of mathematics, open to linguists, mathematicians, and statisticians alike. The initial step would be to develop a grammar of mathematical expressions. Classroom experiments could test whether that grammatical structure is useful for teaching. For statisticians, an explicit grammar is a tool for understanding how we use mathematical objects as models, and talk about the relationship between those objects, data, and the underlying thing we are trying to model. We can leverage such a grammar to help us communicate these relationships to non-experts and students alike.

\newpage
\bibliographystyle{siam}
\bibliography{References}
\end{document}